\documentclass[12 pt]{article}
\usepackage{amsmath,amssymb,euscript,latexsym,graphicx,fullpage}
\usepackage{color,amstext,subfig,parskip,balance}
\graphicspath{{./},{./figures/}}

\newtheorem{thm}{Theorem}

\newtheorem{prop}{Proposition}

\newcommand{\ee}{\end{equation}}
\newcommand{\be}{\begin{equation}}
\newcommand{\la}{\langle}
\newcommand{\ra}{\rangle}

\newcommand{\mR}{{\mathbb R}}
\newcommand{\vv}{{\rm vol}}
\newcommand{\cD}{{\cal D}}
\newcommand{\cE}{{\cal D}}
\newcommand{\cH}{{\cal H}}

\newcommand{\cS}{{{\cal S}}}

\newcommand{\cL}{{\mathcal L}}

\newcommand{\trace}{\operatorname{tr}}

\def\spacingset#1{\def\baselinestretch{#1}\small\normalsize}

\definecolor{grey}{rgb}{0.6,0.6,0.6}
\definecolor{lightgray}{rgb}{0.97,.99,0.99}

\begin{document}
\title{Matrix Optimal Mass Transport:\\a Quantum Mechanical Approach}

\author{Yongxin Chen, Tryphon T. Georgiou, and Allen Tannenbaum
\thanks{Y.\ Chen is with the Department of Medical Physics, Memorial Sloan Kettering Cancer Center, NY; email: chen2468@umn.edu}
\thanks{T.\ T. Georgiou is with the Department of Mechanical and Aerospace Engineering, University of California, Irvine, CA; email: tryphon@uci.edu}
\thanks{A.\ Tannenbaum is with the Departments of Computer Science and Applied Mathematics \& Statistics, Stony Brook University, NY; email: allen.tannenbaum@stonybrook.edu}}

\maketitle

\begin{abstract}
In this paper, we describe a possible generalization of the Wasserstein 2-metric, originally defined on the space of scalar probability densities, to the space of Hermitian matrices with trace one, and to the space of matrix-valued probability densities. Our approach follows a computational fluid dynamical formulation
of the Wasserstein-2 metric and utilizes certain results from the quantum mechanics of open systems, in particular the Lindblad equation. It allows determining the gradient flow for the quantum entropy relative to this matricial Wasserstein metric. This may have implications to some key issues in quantum information theory.
\end{abstract}

\section{Introduction}

Optimal mass transport is a rich area of research with applications to
numerous disciplines including econometrics, fluid dynamics,
automatic control, transportation, statistical physics, shape
optimization, expert systems, and meteorology \cite{Rachev,Villani}.
The original problem was first formulated by the civil engineer Gaspar Monge
in 1781, and concerned finding the optimal way, in the sense of
minimal transportation cost, of moving a pile of soil from one site
to another. Much later the problem was extensively analyzed by
Kantorovich~\cite{Kantorovich1948} with a focus on economic resource
allocation, and so is now known as the Monge--Kantorovich (MK) or
optimal mass transport (OMT) problem.

In this paper, we develop a non-commutative counterpart of optimal transport where density matrices $\rho$ (i.e., Hermitian matrices that are positive-definite and have unit trace) replace probability distributions, and where ``transport'' corresponds to a flow on the space of such matrices that minimizes a corresponding action integral. In some recent work, \cite{lipeng}, a certain approach was formulated that had its basis on Kantorovich's idea of regularization on a joint distribution in a suitable product space. In contrast, in the present work, we employ generalizations of the seminal approach of  Benamou and Brenier \cite{French}. In particular, we utilize ideas from quantum mechanics \cite{Sigal} in a Benamou--Brenier framework, our version of non-commutative optimal mass transport
allows us to define geodesics on the space of positive-densities. An alternative approach to matrix optimal transport that is also based on ideas from quantum mechanics (and especially the Clifford algebra which entered into quantum physics from the Dirac equation) was proposed earlier by Carlen and Maas \cite{Carlen}. We should note that many of the definitions we provide for the gradient and divergence are similar to corresponding concepts in \cite{Carlen}.

The remainder of this paper may be summarized as follows. In Section~\ref{sec:mk}, we describe the Benamou-Brenier approach to the Monge-Kantorovich problem, and then in Section~\ref{sec:scalar} we show how this leads to a Riemannian structure on the space of (scalar) probability densities. Sections~\ref{sec:quantum} and \ref{sec:quantum1} are the key parts of the present work. Here we propose the non-commutative analogue of the Wasserstein 2-metric for matrix probability densities. In Section~\ref{sec:entropy}, we write down the corresponding gradient ascent equation for the entropy (based on the Wasserstein  distance), and finally in Section~\ref{sec:conclusions}, we describe some possible future research directions.

\section{Background}
In this section we highlight important concepts and constructs from the theory of OMT for scalar-valued distributions. We focus in particular on the fluid dynamical formulation of Benamou and Brenier and the Riemannian structure on the space of densities that originates in the work of Otto and his coworkers. This background section is sketchy but will allows us to draw analogies with the matrix-valued counterpart of the theory that will follow. See \cite{Villani} and the references therein for more details of OMT.

\subsection{Fluid dynamical approach to OMT} \label{sec:mk}

Here, the $\rho$'s represent positive distributions (density functions\footnote{More generally, one may consider positive measures but this will be avoided for simplicity and for ease of correspondence with the matrix case.}) on a linear space such as $\mR^n$ (assumed throughout this section). It was shown in \cite{French} that the Monge-Kantorovich problem \cite{Rachev, Villani} with a quadratic cost, i.e., the problem to transfer mass which is initially distributed according to $\rho_0$ to a final target distribution $\rho_1$\footnote{The two marginals $\rho_0$ and $\rho_1$ are assumed to have finite second order moments.}, optimally via a transfer map $x\mapsto T(x)$ that minimizes cost, may be given a computational fluid formulation.

Indeed, the mass transfer cost, which is referred to as the {\em Wasserstein distance} $W_2$ between the two densities $\rho_0$ and $\rho_1$,
\[
W_2(\rho_0,\rho_1)^2 := \inf\{ \int \|x-T(x)\|^2 \rho_0(x)dx\mid \rho_0(x) = \rho_1(T(x))\det(\nabla T(x))\}
\]
can also be expressed as the infimum of the ``action integral''
\be\label{actionintegral}
\inf \int \int_0^1 \rho(t,x) \|v(t,x)\|^2 \, dt \, dx
\ee
over a ``flow'' of time-varying densities $\rho(t,x)$ and velocity fields $v(t,x)\in\mR^n$ (weakly) satisfying the {\bf continuity equation}
\begin{eqnarray}
\frac{\partial \rho}{\partial t} + \nabla \cdot (\rho v) &=& 0,
\label{eqcont}\end{eqnarray}
and the boundary conditions
\begin{equation}
\rho(0, \cdot) = \rho_0, \; \rho(1, \cdot) = \rho_1 \nonumber.
\end{equation}
The problem has an elegant structure!
The optimal solution may be characterized by the follow condition.
\begin{thm}
The solution of the OMT problem \eqref{actionintegral} is
	\begin{subequations}\label{eq:optimalomt}
	\begin{equation}
		v(t,x)=-\nabla \phi(t,x),
	\end{equation}
	where $\phi$ and the corresponding flow $\rho$ satisfy
	\begin{equation}
		\frac{\partial \phi}{\partial t} - \frac{1}{2}\|\nabla \phi\|^2=0,
	\end{equation}
	and
	\begin{equation}
		\frac{\partial \phi}{\partial t}-\nabla\cdot(\rho \nabla \phi)=0.
	\end{equation}
	\end{subequations}
\end{thm}

It turns out that the functional \eqref{actionintegral} can be conveniently expressed as
\be \label{actionintegral2}
\inf \int \int_0^1 \rho(t,x)^{-1} \|u(t,x)\|^2 \, dt \, dx
\ee
with $u(t,x)=\rho(t,x)v(t,x)$ a momentum field,
which is convex the density and momentum pair $(\rho,u)$. Under fairly general conditions, the infimum is attained and the minimizing velocity field $v$ is unique. Moreover, the minimizing velocity field turns out to be the gradient $\nabla \varphi$ of a convex function $\varphi$ and the corresponding flow is simply
$x +t\
(\nabla \varphi(x) -x)$, where $\nabla \varphi(x)=:T(x)$ is precisely the solution to the Monge--Kantorovich problem \cite{Villani}. Thus, the analysis in \cite{French}, with the introduction of the action integral, provides a {\em physically motivated dynamical re-interpretation of the MK problem}.

\subsection{Riemannian manifold structure on scalar probability densities} \label{sec:scalar}

Much more can be gained by intuition that has been provided by the physical insight. Indeed, a starling connection between entropy functionals, the heat equation, and the geometry induced by the Wasserstein distance emerged \cite{Otto}. We now briefly touch upon these as it will allows to draw analogies in the matricial setting that follows.

Consider the manifold of scalar densities on $\mR^n$ integrating to $1$,
\[
\cD := \{\rho \ge 0: \int \rho =1 \}.
\]
The tangent space at a given point $\rho$ may be identified with functions $\delta$ integrating to $0$,
\[
T_\rho \cong \{\delta : \int \delta =0 \}.
\]
The manifold $\cD$ admits a Riemannian type structure that
induces the Wasserstein distance. The key idea essentially originated in Jordan {\em et al.} \cite{Jordan} and was developed into a powerful geometric approach to OMT by Otto in \cite{Otto}; see also  \cite{Villani}.

More specifically, under suitable assumptions on differentiability for $\rho \in \cD$ and $\delta \in T_\rho$, one
solves the Poisson equation
\begin{equation} \label{poisson} \delta = -\nabla \cdot (\rho \nabla g). \end{equation}
This allows identifying elements $\delta$ in the tangent space with functions $g$, up to additive constant;
thus, given $\delta$ we denote the solution of (\ref{poisson}) by $g_\delta$ and the corresponding vector field by $v_\delta:=\nabla g_\delta$.
Then given, $\delta_1, \delta_2 \in T_\rho$, we can define the inner product
\begin{equation} \label{inner}
 \la \delta_1, \delta_2 \ra_\rho := \int \rho \langle v_{\delta_1}, v_{\delta_2}\rangle,
 \end{equation}
where $\la \cdot,\cdot\ra$ denotes the standard inner product on $\mR^n$.
An integration by parts argument, shows that this inner product will
exactly induce the Wasserstein distance $W_2(\rho_0,\rho_1)$ given by
Equation~(\ref{actionintegral}). Thus, given two ``points'', $\rho_0, \rho_1 \in \cD$, the minimizer of the Benamou-Brenier formulation which coincides with the {\em displacement interpolating} curve \cite{mccann} between the two densities, $\rho(t,\cdot)$, is precisely a Wasserstein geodesic. Interestingly, using integration by parts
\begin{eqnarray} \label{norm}
 \|\delta\|_\rho^2=\la \delta, \delta \ra_\rho &=& \int \rho \nabla g_\delta \cdot \nabla g_\delta  = -\int g_\delta \nabla \cdot (\rho \nabla g_\delta) = \int \delta g_\delta .
\end{eqnarray}

Note that the distance between $\rho_0$ and $\rho_1$ may be rewritten as
	\[
	W_2(\rho_0,\rho_1)=\min_\rho \int_0^1 \|\dot\rho(t)\|_{\rho(t)} dt
	=\min_\rho \int_0^1 \sqrt{\la\dot\rho(t),\dot\rho(t)\ra_{\rho(t)}} dt,
	\]
where the minimum is taken over all the piecewise smooth curves connecting $\rho_0$ and $\rho_1$.

\subsection{Gradient flow of the entropy}\label{gradientflowclassical}
We close by sketching the fact that the gradient flow  with respect to the Wasserstein geometry of the entropy functional
\[
S(\rho)=-\int\rho\log\rho,
\]
$\rho\in\cD$, is given by the heat equation (this is due to \cite{Otto}, see also \cite{TGT}). Indeed,
evaluate $S$ along a 1-parameter family in $\cD$, $\rho(t,\cdot)$,
and take the derivative with respect to $t$.  Since $\int \rho =1$,
\begin{equation}
\label{energy}
\frac{dS}{dt} = -\int (\frac{\partial \rho}{\partial t} \log \rho + \frac{\partial \rho}{\partial t}) = -\int (\frac{\partial \rho}{\partial t} \log \rho),
\end{equation}
where $\rho_t$ denotes partial derivative with respect to time. Now
noting the characterization of the Wasserstein norm from Equation~(\ref{norm}), we see that the
the steepest gradient direction (with respect to the Wasserstein metric) is given by $g=-\log\rho$. This gives
$$\frac{\partial \rho}{\partial t} = \nabla\cdot(\rho \nabla \log \rho)= \Delta \rho,$$
which is the \textbf{\emph{linear heat equation}}.

\section{Matricial Wasserstein geometry}

For a range of problems in spectral analysis of vector-valued time series as well as in quantum mechanics, statistics of the underlying experimental setting are ecapsulated in matrix-based models. For instance, in quantum mechanics the statistical description of a system is via a state $\rho$ which is a positive element in a corresponding $C^*$-algebra of operators on a Hilbert space. For us, the Hilbert space will always be finite-dimensional and hence $\rho$ would simply be a Hermitian matrix with trace one. Likewise, in multivariable time series and vector-valued random variables (see e.g.\ \cite{lipeng}), $\rho$ may represent a matrix-valued power spectral density or a covariance. In those cases, the integral of the trace or the trace, respectively, represent power and can be normalized to one for our purposes.

Our aim is to develop a geometric framework that will have bearing on
problems in quantum information theory as well as multivariable time series.
Throughout the rest of the paper, the $\rho$'s represent density matrices
(positive-definite Hermitian matrices of trace one, or suitably normalized positive-definite Hermitian-valued functions), and we develop a non-commutative counterpart of the Wasserstein geometry by building on Quantum Mechanical insights and constructs. The key is to {\em devise a suitable notion of a continuity equation} as well as a {\em matrix-valued counterpart of the Benamou-Brenier action integral}. These are done next.

\subsection{Quantum continuity equation} \label{sec:quantum}

Our approach is based on the \textbf{\emph{Lindblad equation}} which describes the evolution of open quantum systems. These are thought of as coupled to a larger system (the environment, ancilla) and, thereby, cannot in general be described by a wave function. The proper description is in terms of a density operator $\rho$ \cite{Sigal} which in turn obeys Lindblad's equation (in diagonal form)
	\begin{equation}\label{eq:lindblad}
	\dot\rho =-i [H, \rho]+\sum_{k=1}^N( L_k\rho L_k^*-\frac{1}{2}\rho L_k^*L_k-\frac{1}{2}L_k^*L_k\rho),
	\end{equation}
where $^*$ denotes conjugate transpose, and throughout, we assume that $\hbar =1$.
The first term on the right-hand side describes the evolution of the state under the effect of the Hamiltonian $H$, and it is unitary (energy preserving). The other the terms on the right-hand side model diffusion and, thereby, capture the dissipation of energy -- it is the quantum analogue of Laplace's operator $\Delta$. The calculus we develop next actually underscores the parallels.

Regarding notation, we denote by $\cH$ and $\cS$ the set of  $n\times n$ Hermitian and skew-Hermitian matrices, respectively. Since matrices are $n\times n$ throughout, we dispense of $n$ in the notation. We also denote the space of block-column vectors consisting of $N$ elements in $\cS$ and $\cH$ as $\cS^N$, respectively $\cH^N$. We let $\cH_+$ and $\cH_{++}$ denote the cones of nonnegative and positive-definite matrices, respectively, and
	\begin{equation}\label{eq:D}
	\cD_+ :=\{\rho \in \cH_{++} \mid \trace(\rho)=1\}.
	\end{equation}
Clearly, the tangent space of $\cD_+$, at any $\rho\in \cD_+$, is now
	\begin{equation}\label{eq:Trho}
		T_\rho=\{ \delta \in \cH \mid \trace(\delta)=0\}.
	\end{equation}
We also use the standard notion of inner product:
	\[
		\langle X, Y\rangle=\trace(X^*Y),
	\]
for both $\cH$ and $\cS$.
For $X, Y\in \cH^N$ ($\cS^N$),
	\[
		\langle X, Y\rangle=\sum_{k=1}^N \trace(X_k^*Y_k).
	\]
Given $X=[X_1^*,\cdots,X_N^*]^* \in \cH^N$ ($\cS^N$), $Y\in \cH$ ($\cS$), denote
	\[
		XY=\left[\begin{array}{c}
		X_1\\
		\vdots \\
		X_N
		\end{array}\right]Y
		:=
		\left[\begin{array}{c}
		X_1Y\\
		\vdots \\
		X_N Y
		\end{array}\right],
	\]
and
	\[
		YX=Y\left[\begin{array}{c}
		X_1\\
		\vdots \\
		X_N
		\end{array}\right]
		:=
		\left[\begin{array}{c}
		YX_1\\
		\vdots \\
		YX_N
		\end{array}\right].
	\]

Throughout, we make the assumption that $L_k=L_k^*$, i.e., $L_k\in \cH$ for all $k\in 1 \ldots, N$. Under this assumption, we can define
	\begin{equation}\label{eq:gradient}
		\nabla_L: \cH \rightarrow {\cS}^N, ~~X \mapsto
		\left[ \begin{array}{c}
		L_1 X-XL_1\\
		\vdots \\
		L_N X-X L_N
		\end{array}\right]
	\end{equation}
as the \textbf{\emph{gradient operator}}.
Note that $\nabla_L$ acts just like the standard gradient operator and shares many useful properties, such as,
	\be \label{eq:product}
		\nabla_L(XY+YX)=(\nabla_L X) Y+ X (\nabla_L Y)+(\nabla_L Y) X+ Y (\nabla_L X),~~\forall X, Y\in \cH.
    \ee

The dual of $\nabla_L$, which is an analogue of the (negative) \textbf{\emph{divergence operator}}, is given by
	\begin{equation}\label{eq:divergence}
		\nabla_L^*: {\cS}^N \rightarrow \cH,~~Y=
		\left[ \begin{array}{c}
		Y_1\\
		\vdots \\
		Y_N
		\end{array}\right]
		\mapsto
		\sum_k^N L_k Y_k-Y_k L_k.
	\end{equation}
Hence, the duality
	\[
		\langle \nabla_L X, Y\rangle =\langle X, \nabla_L^* Y\rangle
	\]
is straightforward.

With these definitions we can easily calculate the (matricial) \textbf{\emph{Laplacian}} as
	\[
		\Delta_L X:=-\nabla_L^*\nabla_L X=\sum_{k=1}^N( 2L_k X L_k^*-X L_k^*L_k-L_k^*L_k X),~~X\in \cH,
	\]
which is exactly (after scaling by $1/2$) the diffusion term in the Lindblad equation \eqref{eq:lindblad}. Therefore Lindblad's equation (under the assumption that $L_k=L_k^*$) can be rewritten as
	\[
		\dot\rho =-i [H, \rho]+\frac{1}{2} \Delta_L \rho.
	\]
Moreover, using the gradient operator \eqref{eq:gradient} and its adjoint \eqref{eq:divergence}, we can now introduce a corresponding matricial \textbf{\emph{continuity equation}}, and in fact, a family of such equations,
\begin{equation}\label{eq:continuitygen}
	\dot \rho=\nabla_L^*  M_\rho(v), 
	\end{equation}
where $M_\rho(v)$ can be any \textbf{\emph{non-commutative multiplication}} between $\rho$ and $v$ that maps the ``velocity field'' $v\in \cS^N$ to a ``momentum field'' $M_\rho(v)\in \cS^N$.

Usually, in the Lindblad equation \eqref{eq:lindblad}, $N$ is taken to be $n^2-1$. However, in general, we may choose $N\le n^2-1$, as needed, possibly large enough such that in \eqref{eq:continuitygen} we are able to cover the whole tangent space $T_\rho$ at $\rho$ for all $\rho\in\cD_+$.
In particular, we need $\nabla_L$ to have the property that the identity matrix $I$ spans its null space. For instance, one can choose $L_1,\ldots,L_N$ to be a basis of the Hermitian matrices $\cH$, in which case $N=n(n+1)/2$. Obviously this construction ensures that the null space of $\nabla_L$ is spanned by $I$.

We will consider two interesting cases of non-commutative multiplication and the corresponding continuity equation, each of which has its own distinct properties.\footnote{An interesting third case that is not discussed herein is $M_\rho(v):=\rho^{1/2}v\rho^{1/2}$.} The first case will be for
\begin{subequations}
\begin{equation}\label{eq:product1}
		M_\rho(v):=\frac{1}{2}(\rho v+v\rho),
\end{equation}
which gives
\begin{equation}\label{eq:continuity}
		\dot \rho=\frac{1}{2} \nabla_L^* (\rho v+v\rho)
	\end{equation}
\end{subequations}
and $v=[v_1^*,\ldots,v_N^*]^* \in \cS^N$. Clearly $\rho v+v\rho \in \cS^N$, which is consistent with the definition of $\nabla_L^*$. We will refer to this as the \textbf{\emph{anti-commutator}} case, as it is standard to refer to
\[
\rho v+v\rho=:\{\rho,v\}
\]
as the anti-commutator when applied to elements of an associative algebra. The second case will be for the Feynman-Kubo-Mori \cite{feynman,hiai} product
\begin{subequations}\label{eq:2}
\begin{equation}\label{eq:product2}
		M_\rho(v):=\int_0^1 \rho^s v\rho^{1-s} ds,
\end{equation}
which leads to a continuity equation
\begin{equation}\label{eq:continuity2}
		\dot \rho=\nabla_L^* \int_0^1 \rho^s v\rho^{1-s} ds
	\end{equation}
\end{subequations}
that we will refer to as the \textbf{\emph{logarithmic}} case. Here too, $ \int_0^1 \rho^s v\rho^{1-s} ds\in \cS^N$, which is consistent with the definition of $\nabla_L^*$.
The terminology ``logarithmic'' will become clearer in Section~\ref{sec:entcont} below. The analysis of both equations and the resulting Wasserstein metrics is quite similar.  Both give a fluid dynamic formulation of optimal transport on the space $\cD_+$ of density matrices, thereby extending the work of Benamou and Brenier \cite{French}. We will begin with the anti-commutator case and then sketch the logarithmic one, both in the next section.

\subsection{Matricial optimal mass transport}\label{sec:quantum1}

We treat separately the anticommutator and logarithmic cases of the two alternative noncommutative products $M_\rho(v)$ between $\rho$ and $v$. The developments are completely analogous.

\subsubsection{Anti-commutator formulation}\label{sec:antiformula}

Given two density matrices $\rho_0, \rho_1 \in \cD_+$, one can formulate the optimization problem
 	\begin{subequations}\label{eq:quantumomt}
 	\begin{eqnarray}\label{eq:quantumomt1}
	 W_{2,a}(\rho_0, \rho_1)^2:=&&\min_{\rho\in \cD_+, v\in \cS^N} \int_0^1 \trace(\rho v^*v) dt,\\
	&&\dot \rho=\frac{1}{2} \nabla_L^* (\rho v+v\rho), \label{eq:quantumomt2}\\
	&& \rho(0)=\rho_0, ~~\rho(1)=\rho_1,  \label{eq:quantumomt3}
	\end{eqnarray}
	\end{subequations}
and define the (matricial, ``anti-commutator'') \textbf{\emph{Wasserstein distance}} $W_{2,a}(\rho_0, \rho_1)$ between $\rho_0$ and $\rho_1$ to be the square root of the minimum of the cost \eqref{eq:quantumomt1}. Note here we have adopted the notation that, $v^*v=\sum_{k=1}^N v_k^*v_k$ for $v\in \cS^N$.

Let $\lambda(\cdot) \in \cH$ be a smooth Lagrangian multiplier  for the constraints \eqref{eq:quantumomt2} and construct the Lagrangian
	\begin{eqnarray*}
		\cL(\rho,v,\lambda)&=&\int_0^1 \left\{\frac{1}{2}\trace(\rho v^*v)-\trace(\lambda
		(\dot \rho-\frac{1}{2} \nabla_L^* (\rho v+v\rho)))\right\}dt
		\\&=& \int_0^1 \left\{\frac{1}{2}\trace(\rho v^*v)-\frac{1}{2}\trace(\nabla_L\lambda (\rho v+v\rho))+
		\trace(\dot\lambda\rho)\right\}dt -\trace(\lambda(1)\rho_1)+\trace(\lambda(0)\rho_0).
	\end{eqnarray*}
Point-wise minimizing the above over $v$ yields
	\[
		v_{opt}(t)=-\nabla_L \lambda(t).
	\]
The corresponding minimum is
	\[
		\int_0^1 \left\{-\frac{1}{2}\trace(\rho (\nabla_L \lambda)^* (\nabla_L \lambda))+
		\trace(\dot\lambda\rho)\right\}dt -\trace(\lambda(1)\rho_1)+\trace(\lambda(0)\rho_0),
	\]	
from which we conclude the following sufficient conditions for optimality. This optimality condition should be compared with \eqref{eq:optimalomt}.
\begin{thm}\label{thm:matrixomtopt}
Suppose there exists $\lambda(\cdot)\in \cH$ satisfying
	\begin{subequations}
	\begin{equation}
		\dot\lambda=\frac{1}{2}(\nabla_L \lambda)^* (\nabla_L \lambda)
		=\frac{1}{2}\sum_{k=1}^N(\nabla_L \lambda)_k^* (\nabla_L \lambda)_k
	\end{equation}
	such that the solution of
	\begin{equation}
		\dot \rho=-\frac{1}{2} \nabla_L^* (\rho \nabla_L \lambda+\nabla_L \lambda\rho)
	\end{equation}
	\end{subequations}
	matches the marginals $\rho(0)=\rho_0, \rho(1)=\rho_1$, then the pair $(\rho, v)$ with $v=-\nabla_L \lambda$ solves \eqref{eq:quantumomt}.
\end{thm}

The Wasserstein distance function $W_{2,a}$ gives a Riemannian structure
	\[
	\la \delta_1, \delta_2\ra_\rho=\trace(\rho \nabla\lambda_1^*\nabla\lambda_2)
	\]
on the tangent space \eqref{eq:Trho}
\[
T_\rho=\{\delta \in \cH \mid \trace(\delta)=0\}.
\]
Here $\lambda_j,~j=1, 2$ is the solution to the ``Poisson'' equation
	\begin{equation}\label{eq:poisson1}
		\delta_j=-\frac{1}{2}\nabla_L^*(\rho\nabla_L\lambda_j+\nabla_L\lambda_j \rho).
	\end{equation}

The proof of existence and uniqueness of the solution of \eqref{eq:poisson1} follows exactly along the same lines as in \cite[Section 3.2]{Carlen}.
The solution of \eqref{eq:poisson1} can be, in fact, calculated as the unique $\lambda$ (up to the addition of a scaled identity matrix $\alpha I$) such that $\nabla_L\lambda$ satisfies the Lyapunov equation
	\begin{equation}\label{eq:lyap}
		\rho\nabla_L\lambda+\nabla_L\lambda \rho =2 \nabla_L \Delta_L^{-1} \delta.
	\end{equation}
The expression $\Delta_L^{-1} \delta$ makes sense as $\delta$ is orthogonal to $I$, which spans the null space of $\nabla_L$, therefore the null space of $\Delta_L$.
Clearly,
	\[
		-\frac{1}{2}\nabla_L^*(2\nabla_L \Delta_L^{-1} \delta)=\Delta_L \Delta_L^{-1} \delta=\delta,
	\]
which is consistent with \eqref{eq:poisson1}.
Now since $\rho\in \cD_+$, we can pick $\nabla_L\lambda$ so that $i\nabla_L\lambda$ is the unique maximal solution of the Lyapunov equation \eqref{eq:lyap}. More interestingly, given a tangent vector $\delta$, $\nabla_L \lambda$ is the unique minimizer of $\trace(\rho v^*v)$ over all the velocity $v\in \cS^N$ satisfying
	\[
		\delta=-\frac{1}{2}\nabla_L^*(\rho v+v \rho).
	\]

Therefore, with the above definition of inner product, $W_{2,a}(\cdot, \cdot)$ indeed defines a metric on $\cD_+$. Moreover, the distance between two given $\rho_0,\rho_1\in \cD_+$ can be rewritten as
	\[
		W_{2,a}(\rho_0, \rho_1)=\min_\rho \int_0^1 \sqrt{\la \dot\rho(t),\dot\rho(t) \ra_{\rho(t)}}dt,
	\]
where the minimum is taken over all the piecewise smooth path on the manifold $\cD_+$.

The Wasserstein distance $W_{2,a}$ can be extended to the closure of $\cD_+$, i.e., the space (denoted by $\cD$) of all positive semidefinite matrices with trace $1$, by continuity. For any two matrices $\rho_0,\rho_1\in \cD$, we can construct sequences $\{\rho_0^j\},\{\rho_1^j\}$ in $\cD_+$ converging to $\rho_0$ and $\rho_1$, respectively, in Frobenius norm. It can be shown that the definition $W_2(\rho_0,\rho_1):=\lim_{j\rightarrow\infty} W_2(\rho_0^j,\rho_1^j)$ makes sense, see \cite[Proposition 4.5]{Carlen}.

\bigskip \noindent
{\bf Remark 1:}
For computational  purposes, it is important to note that Problem \eqref{eq:quantumomt} can be cast as a convex optimization problem in a manner analogous to that in the scalar case \cite{French}, cf.\ Equation \eqref{actionintegral2}.  Define $u:=\rho v=[u_1^*, \ldots,u_N^*]^*$ and $\bar{u}:=[u_1, \ldots,u_N]^*$, then
	\[
		\trace(\rho v^*v)=\sum_{k=1}^N \trace(\rho v_k^*v_k)=\sum_{k=1}^N \trace((\rho v_k)^* \rho^{-1} \rho v_k)
		=\trace(u^*\rho^{-1}u),
	\]
and we readily arrive at the equivalent convex optimization problem
	 \begin{subequations}\label{eq:quantumomtcvx}
 	\begin{eqnarray}\label{eq:quantumomtcvx1}
	W_{2,a}(\rho_0, \rho_1)^2=&&\min_{\rho, u} \int_0^1 \trace(u^*\rho^{-1}u) dt,\\
	&&\dot \rho=\frac{1}{2} \nabla_L^* (u-\bar u), \label{eq:quantumomtcvx2}\\
	&& \rho(0)=\rho_0, ~~\rho(1)=\rho_1  \label{eq:quantumomtcvx3}.
	\end{eqnarray}
	\end{subequations}

\subsubsection{Transport with spatial component: the anti-commutator case} \label{sec:quantum2}

In applications it is often the case that one has to deal with matrix-valued distributions on dimensions which may represent space or frequency. Thus, in this case, the $\rho$'s may be $\cH_+$-valued functions on $E\subset\mR^m$. For instance, in the context of multivariable time series analysis it is natural to consider $m=1$, see, e.g., \cite{lipeng}.
For simplicity, we assume $E$ to be a (convex) connected compact set.
Therefore, in this section
\begin{equation}
\cD=\{ \rho(\cdot) \mid \rho(x)\in \cH_+ \mbox{ for }x\in E \mbox{ such that }\int_{\mR^m} \trace(\rho(x))dx =1\}.
\end{equation}
Let $\cD_+$ denote the interior of $\cD$, and in order to avoid proliferation of notation we use the same symbol $\cD$ ($\cD_+$) as above.
By combining the standard continuity equation on the Euclidean space and the continuity equation for density matrices \eqref{eq:continuity}, we obtain a continuity equation on $\cD_+$ for the flow $\rho(t,x)$ as
	\begin{equation}\label{eq:continuityspec}
		\frac{\partial \rho}{\partial t}+\frac{1}{2}\nabla_x\cdot(\rho w+w\rho)-\frac{1}{2} \nabla_L^* (\rho v+v\rho)=0.
	\end{equation}
Here $\nabla_x \cdot$ is the standard (negative) divergence operator on $\mR^m$, $w(t,x)\in {\cH}^m$ is the velocity field along the space dimension, and $v(t,x)\in \cS^N$ is the quantum velocity as before.

A dynamic formulation of matrix-valued optimal mass transport between
two given marginals $\rho_0, \rho_1 \in \cD_+$ ensues, namely,
	 \begin{subequations}\label{eq:matrixomt}
 	\begin{eqnarray}\label{eq:matrixomt1}
	W_{2,a}(\rho_0, \rho_1)^2:=&&\min_{\rho\in \cD_+, w\in {\cH}^m, v\in \cS^N} \int_0^1\int_{\mR^m}
	\left \{\trace(\rho w^*w)+\gamma\trace(\rho v^*v)\right\}dxdt\\
	&&\frac{\partial \rho}{\partial t}+\frac{1}{2}\nabla_x\cdot(\rho w+w\rho)-\frac{1}{2} \nabla_L^* (\rho v+v\rho)=0, \label{eq:matrixomt2}\\
	&& \rho(0,\cdot)=\rho_0, ~~\rho(1,\cdot)=\rho_1  \label{eq:matrixomt3}.
	\end{eqnarray}
	\end{subequations}
The coefficient $\gamma>0$ is arbitrary and weighs in the relative significance of the two velocity fields. It is anticipated that, in applications, a suitable choice of $\gamma$ will provide appropriate flows that reflect the underlying physics (trading off the two alternative mechanisms for transfering mass, i.e., via ``flow along $x$'' or via the available ``non-commutative flow''). Once again, we are in a position to define a Wasserstein distance $W_{2,a}(\rho_0, \rho_1)$ between $\rho_0$ and $\rho_1$ via \eqref{eq:matrixomt1}.

A sufficient condition for optimality can be obtained in a similar manner as before.
Here, we let $\lambda(\cdot,\cdot)\in \cH$ be a smooth function and define the Lagrangian
	\begin{align*}
		\cL(\rho,v,w, \lambda)&=
		\int_0^1\int_{\mR^m}\left \{  \frac{1}{2}\trace(\rho w^*w)+\frac{\gamma}{2}\trace(\rho v^*v)\right.\\
		&\left.
		-\trace(\lambda(\frac{\partial \rho}{\partial t}+\frac{1}{2}\nabla_x\cdot(\rho w+w\rho)-\frac{1}{2} \nabla_L^* (\rho v+v\rho)))\right\}dxdt.
	\end{align*}
Integration by parts yields
	\[
	\int_0^1\int_{\mR^m}\left \{  \frac{1}{2}\trace(\rho w^*w)+\frac{\gamma}{2}\trace(\rho v^*v)
		+\trace(\frac{\partial \lambda}{\partial t}\rho)+\frac{1}{2}\langle\nabla_x\lambda,\rho w+w\rho\rangle+\frac{1}{2}
		\langle \nabla_L\lambda, \rho v+v\rho\rangle\right\}dxdt
	\]
Here we have discarded the terms on $\rho_0, \rho_1$. Minimizing the above pointwise over $w, v$ gives expressions for the optimal values as
	\[
		w_{opt}(t,x)=-\nabla_x \lambda(t,x)
	\]
and 	
	\[
		v_{opt}(t,x)=-\frac{1}{\gamma}\nabla_L \lambda(t,x).
	\]
Substituting these back to the Lagrangian we obtain
	\[
		\int_0^1\int_{\mR^m}\left \{  -\frac{1}{2}\trace(\rho (\nabla_x\lambda)^*(\nabla_x\lambda))
		-\frac{1}{2\gamma}\trace(\rho (\nabla_L \lambda)^*(\nabla_L\lambda))
		+\trace(\rho\frac{\partial \lambda}{\partial t})\right\}dxdt,
	\]	
and the sufficient conditions for optimality given below follow.
\begin{thm}
Suppose there exists smooth $\lambda(\cdot,\cdot)\in \cH$ satisfying
	\begin{subequations}
	\begin{equation}
		\frac{\partial\lambda}{\partial t}-\frac{1}{2}(\nabla_x\lambda)^*(\nabla_x\lambda)
		-\frac{1}{2\gamma}(\nabla_L \lambda)^* (\nabla_L \lambda)=0
	\end{equation}
such that the solution of
	\begin{equation}
		\frac{\partial \rho}{\partial t}-\frac{1}{2}\nabla_x\cdot(\rho \nabla_x \lambda+\nabla_x \lambda\rho)+
		\frac{1}{2\gamma} \nabla_L^* (\rho\nabla_L \lambda+\nabla_L \lambda\rho)=0
	\end{equation}
	\end{subequations}
	matches the two marginals $\rho(0,\cdot)=\rho_0, \rho(1,\cdot)=\rho_1.$ Then $(\rho, w=-\nabla_x \lambda, v=-\frac{1}{\gamma}\nabla_L \lambda)$ solves \eqref{eq:matrixomt}.
\end{thm}

The Wasserstein distance $W_{2,a}(\rho,\rho+\delta)$ defines a Riemannian type structure on the tangent space of $\cD_+$ at $\rho$. Given any two tangent vector $\delta_1, \delta_2$ at $\rho$, we can associate them with $\lambda_1, \lambda_2$ by solving the ``Poisson'' equations
	\begin{equation}\label{eq:poisson2}
		\delta_j=\frac{1}{2}\nabla_x\cdot(\rho \nabla_x \lambda_j+\nabla_x \lambda_j\rho)-
		\frac{1}{2\gamma} \nabla_L^* (\rho\nabla_L \lambda_j+\nabla_L \lambda_j\rho), ~~j=1,2.
	\end{equation}
Similar to the argument we had before the the case \eqref{eq:poisson1} without spacial component, the above Poisson equation \eqref{eq:poisson2} has an unique solution. The proof relies on the fact that the null space of the gradient operator
	\[
		\nabla_{L,x}=\left[\begin{array}{c}\nabla_L\\\nabla_x \end{array}\right]
	\]
is spanned by the constant matrix function $I$.

The Riemannian metric can then be defined as
	\[
		\la \delta_1,\delta_2\ra_{\rho}=\trace(\rho\nabla_x\lambda^*\nabla_x\lambda)
		+\frac{1}{\gamma}\trace(\rho\nabla_L \lambda^* \nabla_L \lambda).
	\]
Therefore $W_{2,a}(\cdot,\cdot)$ is a metric on $\cD_+$, and can be rewritten as
	\[
		W_{2,a}=\min_\rho \int_0^1 \sqrt{\left\la\frac{\partial \rho}{\partial t}, \frac{\partial \rho}{\partial t}\right\ra_{\rho(t)}}dt.
	\]
Here the integral is minimized over all the piecewise smooth curves in $\cD_+$ connecting $\rho_0$ and $\rho_1$. As in Section \ref{sec:antiformula}, the Wasserstein distance $W_{2,a}$ can be extended to the closure $\cD$ of $\cD_+$ by continuity.

\bigskip \noindent
{\bf Remark 2:}
As noted earlier, \eqref{eq:matrixomt} can again be cast as a convex optimization problem: define $q=\rho w$, to obtain the equivalent convex problem
	 \begin{subequations}\label{eq:matrixomtcvx}
 	\begin{eqnarray}\label{eq:matrixomtcvx1}
	&&\min_{\rho, q, u} \int_0^1\int_{\mR^m}
	\left \{\trace(q^*\rho^{-1}q)+\gamma\trace(u^*\rho^{-1}u)\right\}dxdt\\
	&&\frac{\partial \rho}{\partial t}+\frac{1}{2}\nabla_x\cdot(q+\bar{q})-\frac{1}{2} \nabla_L^* (u-\bar{u})=0, \label{eq:matrixomtcvx2}\\
	&& \rho(0,\cdot)=\rho_0, ~~\rho(1,\cdot)=\rho_1  \label{eq:matrixomtcvx3}.
	\end{eqnarray}
	\end{subequations}
	
\subsubsection{The logarithmic case} 

We now briefly discuss the case where the non-commutative multiplication of $\rho$ and $v$ is taken to be \eqref{eq:product2}:
\be \nonumber
M_\rho(v)=\int_0^1 \rho^s v\rho^{1-s} ds.
\ee
For the purposes of defining a corresponding Wasserstein geometry we proceed in a manner entirely analogous to that for the anti-commutator case. Hence, we only highlight the key elements.

The corresponding Wasserstein metric between $\rho_0, \rho_1\in \cD_+$ is obtained via	
\begin{subequations}\label{eq:newquantumomt}
 	\begin{eqnarray}\label{eq:newquantumomt1}
	W_{2,b}(\rho_0, \rho_1)^2:=&&\min_{\rho\in \cD_+, v\in \cS^N} \int_0^1\int_0^1 \trace(v^*\rho^s v\rho^{1-s})ds dt\\
	&&\dot \rho=\nabla_L^*\int_0^1 \rho^s v\rho^{1-s} ds, \label{eq:newquantumomt2}\\
	&& \rho(0)=\rho_0, ~~\rho(1)=\rho_1.  \label{eq:newquantumomt3}
	\end{eqnarray}
	\end{subequations}

Employing a similar argument as in Theorem \ref{thm:matrixomtopt} (see also \cite[Theorem 5.3]{Carlen}), we establish the following optimality condition for \eqref{eq:newquantumomt}.
\begin{prop}
Suppose there exists $\lambda(\cdot)\in \cH$ satisfying
	\begin{subequations}\label{eq:optimallog}
	\begin{equation}
		\dot\lambda=\int_0^1\int_0^1\int_0^\alpha\left\{ \frac{\rho^{\alpha-\beta}}{(1-s)I+s\rho}
		(\nabla_L \lambda)^* \rho^{1-\alpha} \nabla_L \lambda\frac{\rho^{\beta}}{(1-s)I+s\rho}\right\}d\beta d\alpha ds
	\end{equation}
	such that the solution of
	\begin{equation}
		\dot \rho=-\nabla_L^* \int_0^1 \rho^s \nabla_L \lambda\rho^{1-s}ds
	\end{equation}
	\end{subequations}
	matches the marginals $\rho(0)=\rho_0, \rho(1)=\rho_1$, then the pair $(\rho, v)$ with $v=-\nabla_L \lambda$ solves \eqref{eq:newquantumomt}.
\end{prop}
The above optimality condition should be compared with \eqref{eq:optimalomt} in the scalar case, and Theorem \ref{thm:matrixomtopt} in the anti-commutator case. Unlike the other two, where $\rho$ doesn't affect $\lambda$ directly, here the two differential equations \eqref{eq:optimallog} are coupled in both directions.

Similarly, for matrix-valued densities, the corresponding metric is obtained via
	 \begin{subequations}\label{eq:newmatrixomt}
 	\begin{eqnarray*}\label{eq:newmatrixomt1}
	W_{2,b}(\rho_0, \rho_1)^2:=&&\min_{\rho\in \cD_+, w\in {\cH}^m, v\in \cS^N} \int_0^1\int_{\mR^m}\int_0^1
	\left \{\trace(w^*\rho^s w\rho^{1-s})+\gamma\trace(v^*\rho^s v\rho^{1-s})\right\}dsdxdt\\
	&&\frac{\partial \rho}{\partial t}+\nabla_x\cdot(\int_0^1\rho^s w\rho^{1-s} ds)-
	\nabla_L^* (\int_0^1\rho^s v\rho^{1-s} ds)=0, \label{eq:newmatrixomt2}\\
	&& \rho(0,\cdot)=\rho_0, ~~\rho(1,\cdot)=\rho_1  \label{eq:newmatrixomt3}.
	\end{eqnarray*}
	\end{subequations}

\section{Gradient flow of the entropy} \label{sec:entropy}

We close by presenting the matricial counterpart of the classical result of \cite{Otto} for the case of scalar-valued distributions that {\em the gradient flow of the entropy is the heat equation} (see Section \ref{gradientflowclassical}). Thus, below, we derive gradient flows for the entropy functional on density matrices with respect to the two alternative Wasserstein geometries.

\subsection{The anticommutator case}

The entropy of density matrices is defined by
	\[
		S(\rho)=-\trace(\rho\log\rho).
	\]
The gradient with respect to $W_{2,a}$ may be calculated as follows. For a given flow $\rho(\cdot)$,
	\begin{eqnarray*}
	\frac{dS(\rho(t))}{dt} &=& -\trace((\log \rho+I)\dot \rho)
	\\&=& -\trace((\log\rho+I)\frac{1}{2}\nabla_L^* (\rho v+v\rho))
	\\&=& -\frac{1}{2}\trace((\nabla_L \log\rho)^*(\rho v+v\rho))
	\\&=& -\trace(\rho v^*\nabla_L \log\rho),
	\end{eqnarray*}
in view of the definition of $W_{2,a}$, we conclude the steepest ascent direction is
	\[
		v=-\nabla_L \log\rho.
	\]
Substituting back to the continuity equation (\ref{eq:continuity}), we obtain the gradient flow
	\be \label{eq:gradient1}
		\dot \rho=-\frac{1}{2} \nabla_L^* (\rho \nabla_L \log\rho+\nabla_L \log\rho \cdot \rho)=-\frac{1}{2}\nabla_L^*(\{\rho, \nabla_L \log \rho\}),
	\ee
where $\{\cdot, \cdot\}$ denotes the anti-commutator as before.

Similarly, we may consider entropy function for matrix-valued densities
	\[
		S(\rho)=-\int_{\mR^m} \trace(\rho \log \rho) dx
	\]
for $\rho\in \cE$ and the associated gradient flow with respect to $W_{2,a}$. The total derivative of $S$ over a flow $\rho(t,\cdot)$ is
	\begin{eqnarray*}
	\frac{dS(\rho(t,\cdot))}{dt} &=& -\int_{\mR^m} \trace((\log \rho+I)\frac{\partial \rho}{\partial t}) dx
	\\&=& -\int_{\mR^m} \trace((\log \rho+I)(-\frac{1}{2}\nabla_x\cdot(\rho w+w\rho)+\frac{1}{2} \nabla_L^* (\rho v+v\rho)))dx
	\\&=& -\frac{1}{2}\int_{\mR^m} \left\{\trace(\nabla_x\log \rho(\rho w+w\rho))-\trace(\nabla_L\log \rho
	(\rho v+v\rho))\right\}dx
	\\&=& \int_{\mR^m} \left\{-\trace(\rho w^*\nabla_x\log \rho)-\trace(\rho v^*\nabla_L\log \rho)\right\}dx,
	\end{eqnarray*}
which indicates, in view of \eqref{eq:matrixomt}, that the steepest ascent direction is
	\[
		w=-\nabla_x\log \rho, ~~~v=-\frac{1}{\gamma}\nabla_L\log \rho.
	\]
Therefore, the gradient flow is now given by
	\begin{eqnarray*}
		\frac{\partial \rho}{\partial t}&=& \frac{1}{2}\nabla_x\cdot(\rho \nabla_x \log\rho+\nabla_x \log\rho \cdot \rho)-
		\frac{1}{2\gamma} \nabla_L^* (\rho\nabla_L \log\rho+\nabla_L \log\rho \cdot \rho)\\
&=&\frac{1}{2} \nabla_x\cdot (\{\rho, \nabla_x \log \rho\})-\frac{1}{2\gamma}\nabla_L(\{\rho, \nabla_L \log \rho\}) .
	\end{eqnarray*}

\bigskip \noindent
{\bf Remark 3:}
Note that in both of the above cases, the gradient flow of the entropy is nonlinear, which should be contrasted with the linear heat equation that arises in the scalar case (as noted in Section \ref{gradientflowclassical}
following \cite{Jordan,Otto}).
Indeed, Equation~(\ref{eq:gradient1}) is a second order nonlinear equation, which is quite different from the linear Linblad equation, and gives the direction of maximal dissipation of quantum information relative to the Wasserstein metric $W_{2,a}$ defined above.

\subsection{The logarithmic case}\label{sec:entcont}
In this section, we will see that in the logarithmic case,
i.e., when using the noncommutative multiplication and corresponding continuity equation in \eqref{eq:2}, the gradient flow with respect to the corresponding Wasserstein geometry of the matricial entropy now gives the quantum-version of the heat equation
$\dot \rho=\Delta_L\rho$, which is the ``dissipative part'' of the Linblad equation.

A key property of, and our choice of the terminology ``logarithmic'' for,
the noncommutative multiplication
\[
M_\rho(v)=\int_0^1 \rho^s v\rho^{1-s} ds,
\]
can be traced in the rather {\em remarkable identity} (see \cite{Carlen} for details)
\be \label{eq:remark}  \nabla_L\rho = \int_0^1 \rho^{s} (\nabla_L \log \rho) \rho^{1-s} ds.
\ee
It represents a logarithmic averaging. Although at first surprising, the identity itself may be readily proven using the product rule, Equation~(\ref{eq:product}), and the fact that
$$\rho = \lim_{j \rightarrow \infty} ( I + \frac{1}{j} \log \rho )^j.$$ The relation (\ref{eq:remark}) works just as well for more general gradient operators
\be \label{eq:remark1}
\nabla_x\rho = \int_0^1 \rho^s (\nabla_x \log\rho ) \rho^{1-s} ds.
\ee

With this in mind, we move onto the gradient flow of the entropy $S(\rho)$ with respect to $W_{2,b}$. For the case where $\rho(t)\in\cD_+$, i.e., $\rho$ does not depend on spacial coordinates, taking the total derivative of $S(\rho)$ over a flow $\rho(\cdot)$ gives
	\begin{eqnarray*}
	\frac{d S(\rho(t))}{dt} &=& -\trace((\log \rho+I)\dot \rho)
	\\&=& -\trace((\log\rho+I)\nabla_L^* \int_0^1 \rho^s v\rho^{1-s} ds)
	\\&=& -\trace((\nabla_L \log\rho)^*\int_0^1 \rho^s v\rho^{1-s} ds),
	\end{eqnarray*}
which points to the greatest ascent direction
	\[
		v=-\nabla_L\log \rho.
	\]	
Now using Equation~(\ref{eq:remark}) we obtain
\be \label{eq:heat}
\dot \rho=-\nabla_L^*\int_0^1 \rho^s (\nabla_L \log\rho) \rho^{1-s} ds=-\nabla_L^*\nabla_L\rho=\Delta_L\rho,
\ee
which is a {\em linear heat equation}, just as in the scalar case.

Similarly, employing Equation~(\ref{eq:remark1}), we see that (arguing as above) that the gradient flow of $S(\rho)$ with respect to $W_{2,b}$ with spatial dimension is
\be \label{eq:heat1}
\frac{\partial \rho}{\partial t}=\Delta_L\rho+\Delta_x\rho.
\ee

\bigskip \noindent
{\bf Remark 4:}
First and foremost, equations~(\ref{eq:heat}-\ref{eq:heat1}) are indeed quite intriguing because of their similarity to the scalar case. But, more importantly, (\ref{eq:heat}) implies that the dissipation part of the Lindblad equation gives a direction in which quantum information (negative of quantum entropy) is decreasing as rapidly as possible with respect to the specific Wasserstein-$W_{2,b}$ geometry on the space of density matrices.

\bigskip \noindent
{\bf Remark 5:}
It should be noted that both heat equations \eqref{eq:gradient1} and \eqref{eq:heat} can be written in the form
	\[
		\dot\rho=-\nabla_L^* M_\rho (\nabla_L \log\rho),
	\]
but with different non-commutative multiplications. This formula of gradient flow of the entropy $S$ even holds for the cases of other more general non-commutative multiplications. The fact that the heat equation becomes linear in the logarithmic case is due to the remarkable relation \eqref{eq:remark}.

\section{Conclusions and further research} \label{sec:conclusions}

In this note, we proposed a possible extension of the Benamou-Brenier approach to optimal mass transport to the non-commutative case of probability density matrices using ideas from quantum theory.
We discussed two cases where the non-commutative multiplication are taken to be anti-commutator and logarithmic mean, respectively. Each of them have certain advantages relative to one another. In the anti-commutator case, the problem can be formulated as a convex optimization problem, and the optimality condition resembles the one in the scalar setting, while in the logarithmic mean case, the linear heat equation is the gradient flow of the entropy.

In the future,  we plan to consider the implications of this Wasserstein metric to problems in quantum information and networks. In particular, one is drawn to explore the implications of Equation~(\ref{eq:heat}) in this regard. Another related direction is the extension of results of Lott and Villani \cite{LV} to this framework. We briefly recall their result.
Let $X$ denote a Riemannian manifold and set
\begin{eqnarray}
{\cal P}^*(X)&:=& \{ \rho \ge 0: \int_X \rho \, d\vv =1\},\\
{\cal P}(X) &:=& \{ \rho \in {\cal P}^*(X): \lim_{\epsilon \searrow 0} \int_{\rho \ge \epsilon} \rho \log \rho \, d\vv< \infty\}.\end{eqnarray}
We define \be \label{Bent} S(\rho) := -\lim_{\epsilon \searrow 0} \int_{\rho \ge \epsilon} \rho \log \rho \, d\vv,
\mbox{ for } \rho \in {\cal P}(X).\ee  Here $\vv$ denotes the usual volume form on $X$.
In \cite{LV}, it is proven that the Ricci curvature bounded from below by $k$ if and only
for every $\rho_0, \rho_1 \in {\cal P}(X),$ there exists a constant speed geodesic $\rho(t)$ with respect to the Wasserstein 2-metric connecting $\rho_0$ and
$\rho_1$ such that
\be \label{eq:entropy.curvature} S(\rho(t)) \ge t S(\rho_0) + (1-t) S(\rho_1)+\frac{kt(1-t)}{2} W_2(\rho_0, \rho_1)^2, \quad 0\le t\le 1.\ee
We would like an analogous result in our framework, with a suitable Hessian replacing the Ricci curvature. See the closely related result in \cite{Carlen} (Proposition 5.11) as well.

\section*{Acknowledgements}

This project was supported by grants from Air Force Office of Scientific Research (AFOSR), NSF, NIH, and a postdoctoral fellowship through Memorial Sloan Kettering Cancer Center. We would like to thank Drs.~Kaoru Yamamoto, Lipeng Ning, and Sei Zhen Khong for some very helpful conversations about matrix optimal mass transport.

\spacingset{.97}

\bibliographystyle{plain}

\begin{thebibliography}{99}

\bibitem{Carlen}
E. Carlen and J. Maas, ``An analog of the 2-Wasserstein metric in non-commutative probability under which the fermionic Fokker-Planck equation is gradient flow for the entropy,''
{\em Commun. Math. Phys.} 331, 887-926 (2014).

\bibitem{feynman}
R. P. Feynman, ``An operator calculus having applications in quantum
electrodynamics,'' {\em Phys. Rev.}, vol. 84, no. 1, pp. 108-128, 1951.

\bibitem{French}
J.-D.~ Benamou and Y.~Brenier,
``A computational fluid mechanics solution to the
Monge--Kantorovich mass transfer problem,''
{\em Numerische Mathematik} {\bf 84} (2000), pp. 375-393.

\bibitem{Sigal}
S. Gustafson and I.~M.~ Sigal, {\em Mathematical Concepts of Quantum Mechanics}, Springer, New York, 2011.

\bibitem{hiai} F. Hiai, D. Petz, G.Toth, ``Curvature in the geometry of canonical correlation,'' {\em Studia Sci. Math. Hungar.} 32, 235-249.

\bibitem{Jordan}
R. Jordan, D. Kinderlehrer, and F. Otto, ``The variational formulation of the Fokker-Planck equation,''
{\em SIAM J. Math. Anal.} {\bf 29} (1998), pp. 1-17.

\bibitem{Kantorovich1948}
L.~V.~Kantorovich, ``On a problem of Monge,'' {\em Uspekhi Mat.
Nauk.} {\bf 3} (1948), pp. 225--226.

\bibitem{mccann} McCann, Robert J. "A convexity principle for interacting gases." {\em Advances in mathematics 128.1} (1997): 153-179.

\bibitem{lipeng}
L.~Ning, T.~Georgiou, and A.~Tannenbaum, ``On matrix--valued Monge–-Kantorovich optimal mass transport,'' {\em IEEE Transactions on Automatic Control} {\bf 60:2} (2015), pp. 373-382.

\bibitem{LV} J. Lott and C. Villani, ``Ricci curvature for metric-measure spaces via optimal transport," {\em Annals of Mathematics,} {\bf 169} (2009) pp. 903-991.

\bibitem{Otto}
F. Otto, ``The geometry of dissipative evolution equations: the porous medium equation,'' {\em Communications in Partial Differential Equations} {\bf 26} (2001), pp. 101-174.

\bibitem{Rachev}
S.~Rachev and L.~R\"uschendorf,
{\em Mass Transportation Problems}, Volumes I and II, Probability
and Its Applications, Springer, New York, 1998.

\bibitem{TGT} E.~Tannenbaum, T.~Georgiou, and A.~Tannenbaum, A., ``Signals and control aspects of optimal mass transport and the Boltzmann entropy,'' in 49th IEEE Conference on Decision and Control (CDC), December 2010.

\bibitem{Villani} C. Villani, {\em Topics in Optimal Transportation,} Graduate Studies in Mathematics, vol.~58, AMS,
Providence, RI, 2003.

\end{thebibliography}

\end{document}